\documentclass{eas}
\usepackage{graphicx}
\newcommand{\kms}{km\,s$^{-1}$}
\begin{document}

\title{Stokes IQUV mapping of $\alpha^2$ CVn \& other Ap stars using ESPaDOnS and NARVAL} 
\runningtitle{MDI of Ap Stars}
\author{J. Silvester}\address{Department of Physics, Engineering Physics \& Astronomy, Queen's University, Kingston, Ontario, Canada, K7L 3N6}
\author{O.Kochukhov}\address{Department of Astronomy and Space Physics, Uppsala University, 751 20, Uppsala, Sweden.}
\author{G.A. Wade}\address{Department of Physics, Royal Military College of Canada, P.O. Box 17000, Station `Forces', Kingston, Ontario, Canada, K7K 7B4}
\begin{abstract}
New spectral line polarisation observations of 7 bright Ap stars have been obtained with the ESPaDOnS and Narval high resolution spectropolarimeters (Silvester et al. 2012). The aim of this data set is produce a series of surface magnetic field and surface chemistry maps for these Ap stars. We present new magnetic maps for the Ap star $\alpha^2$ CVn using these new data and the MDI inversion code INVERS10.
$\alpha^2$ CVn is the first Ap star to be observed during two separate epochs using high resolution phase resolved spectropolarimetric $IQUV$ observations and as such allows us an insight into how stable the surface magnetic structure is over a decade timescale. We show that the new maps give a magnetic field structure consistent with the previous maps obtained by Kochukhov and Wade (2010) from lower quality MuSiCoS spectra taken a decade ago and that the field topology cannot be described by a dipolar or quadrupolar field. 
\end{abstract}
\maketitle
\section{Introduction}
The bright Ap star $\alpha^2$~CVn has been the subject of many observations over the past century, to name but a few; Babcock \& Burd (1952); Pyper (1969); Borra \& Landstreet (1977), Borra \& Vaughan (1978) and one of the most heavily studied magnetic chemically peculiar stars, with the first period determination as early as Farnsworth (1932).   It was not until Kochukhov et al. (2002) employed a new Magnetic Doppler Imaging technique (MDI) (described by Piskunov \& Kochukhov (2002) and Kochukhov \& Piskunov (2002)) that the first high resolution maps of the surface vector magnetic field using Stokes $IV$ observations were made for $\alpha^2$~CVn. These maps were later refined by using linear polarisation profiles (Stokes $Q$ and $U$) in combination with Stokes $IV$ (Kochukhov \& Wade 2010). These maps (along with maps of 53 Cam, Kochukhov et al. 2004) were distinguished from earlier models in that they were computed directly from the observed polarised line profiles, making no {\em a priori} assumptions regarding the large-scale or small-scale topology of the field. The MDI surface magnetic field maps of both stars revealed that their magnetic topologies depart significantly from low-order multipoles. 

These original maps were limited by the observational data: with MuSiCoS being a relatively inefficient medium-resolution spectropolarimeter mounted on a two-metre telescope, only a very small number of lines could be studied.  With the new observations of Ap stars in all four Stokes parameters (Stokes $IQUV$) using ESPaDOnS and Narval as described by Silvester et al. (2012) it is possible to not only study a larger sample of Ap stars, this new higher quality data allows the study of $\alpha^2$~CVn at a resolution not previously possible.  New higher resolution, higher signal to noise data allows us to probe more subtle spectral features which have been unresolved or buried in the noise in the MuSiCoS observations. $\alpha^2$~CVn made an ideal candidate as the first star to map the magnetic topology using the new ESPaDOnS and Narval data and the INVERS10 code, because of the existence of the magnetic topology maps produced using MuSiCoS data. Having this previous map set  allows us to confirm that the new data are compatible with the old data and that newly modelled magnetic field is consistent with the field derived from the MuSiCoS maps. 

\section{Improvement in Data Quality}
As is discussed by Silvester et al. (2012), the new observations obtained with ESPaDOnS and Narval were of much superior quality. The ESPaDOnS and Narval spectra offer a higher resolution than the MuSiCoS spectra (with $R = \lambda / \Delta \lambda \simeq 65000$), and a larger wavelength coverage from 369-1048 nm (with gaps at 922.4 to 923.4 nm, 960.8 to 963.6 nm and 1002.6 to 1007.4 nm) and a much higher signal-to-noise ratio. (The median signal-to-noise ratio of the reduced observations is over 700 per 1.8~\kms\ pixel).  An illustration of the improvement in data quality is given in Fig. 1 which shows a comparison between the  Stokes $Q$ and $U$ profiles of HD 32633 in the Fe II 5018   \AA\ line between ESPaDOnS/Narval and MuSiCoS, with the new data showing much clearer signatures and greatly reduced noise. At the same time Silvester et al. (2012) showed that the new data were consistent with the MuSiCoS data of Wade et al. (2000).

\begin{figure*}
\begin{center}
   \includegraphics[width=0.95\textwidth]{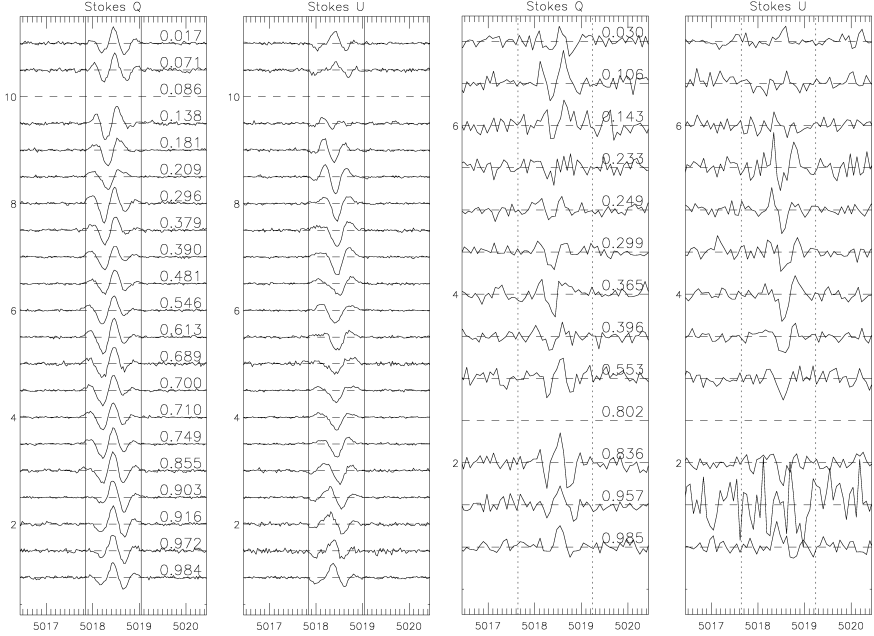}
  \caption{Comparison between Stokes $Q$ and $U$ profiles of HD 32633 in the Fe II 5018   \AA\ line for ESPaDOnS/Narval on the left and MuSiCoS on the right. Rotation phases for each observation are indicated.}
\label{emcomp}
\end{center}
\end{figure*} 

\section{Magnetic field Comparison with MuSiCoS Maps} 
\begin{figure*}
\begin{center}
  \includegraphics[width=0.95\textwidth]{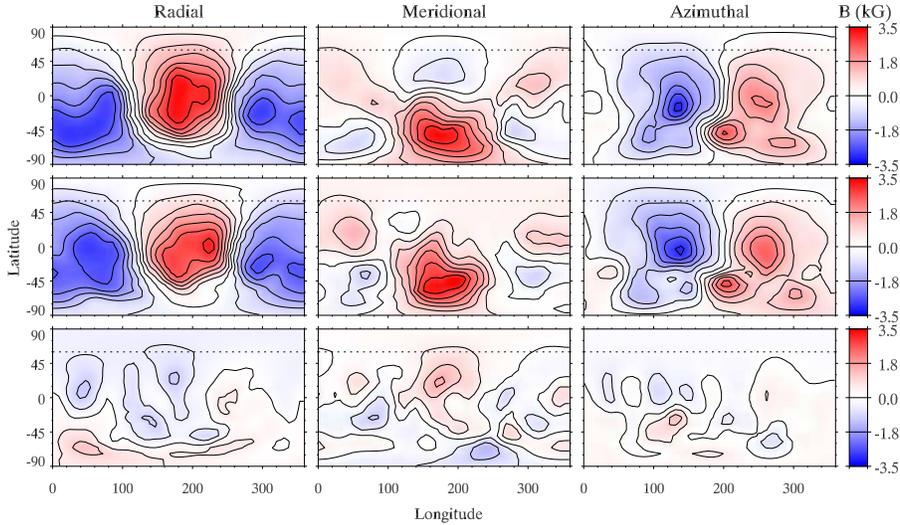}
   \caption{Rectangular maps of three magnetic field vector components for:  Musicos spectra using Invers10 (Kochukhov et al. 2010) (top), best-fit map derived from Espadons/Narval data using INVERS10 (middle) and the difference between the two (bottom).}
\label{field-maps-rec}
\end{center}
\end{figure*}

To allow direct comparison with the original MuSiCoS maps of Kochukhov \& Wade (2010), we reconstructed the magnetic field topology of  $\alpha^2$~CVn using the strong iron lines of Fe II 4923 \AA\  and Fe II 5018 \AA\ using the new ESPaDOnS/Narval data. The MDI mapping is performed using the INVERS10 code (as described by Piskunov \& Kochukhov (2002), Kochukhov \& Piskunov (2002) and Kochukhov \& Wade (2010)).  INVERS10 constructs model line profiles based on an assumed initial surface distribution of free parameters (element abundance and magnetic field geometry) and then iteratively adjusts the parameters until the computed line profiles are in agreement with the observations. A direct comparison of the radial, meridional and azimuthal fields between the two data sets is shown in Fig. \ref{field-maps-rec} in a rectangular representation. Good agreement can be seen between the maps reconstructed from the two data sets for all three field components, with the difference plot showing little structure. This indicates that the two maps are consistent and that the magnetic field is stable over this time period of 10 years spanned by the two data sets. 

\section{Comparison to multipolar geometries}
As has been found in previous mapping of  $\alpha^2$~CVn (Kochukhov \& Wade 2010), the magnetic field topology show complex substructure which could not be described by a dipolar or quadrupolar geometries.  With these new data it is important to investigate whether that this is still the case and confirm that the observed profiles cannot be correctly fitted with a more simple field topology.  To confirm that this is indeed the case,  the profiles for a set of 5 strong and weak iron lines were compared to model profiles one would obtain for a dipolar and dipolar plus quadrupolar geometry.  To accomplish this comparison we have fitted four Stokes parameter observations with a modified version of INVERS10 in which a direct description of the three field components was substituted with a multipolar parameterization similar to the one described by Donati et al. (2006). Further details about our implementation of MDI with multipolar expansion are provided by Kochukhov et al. (2013). 

\begin{figure*}
\begin{center}
 \includegraphics[width=0.95\textwidth]{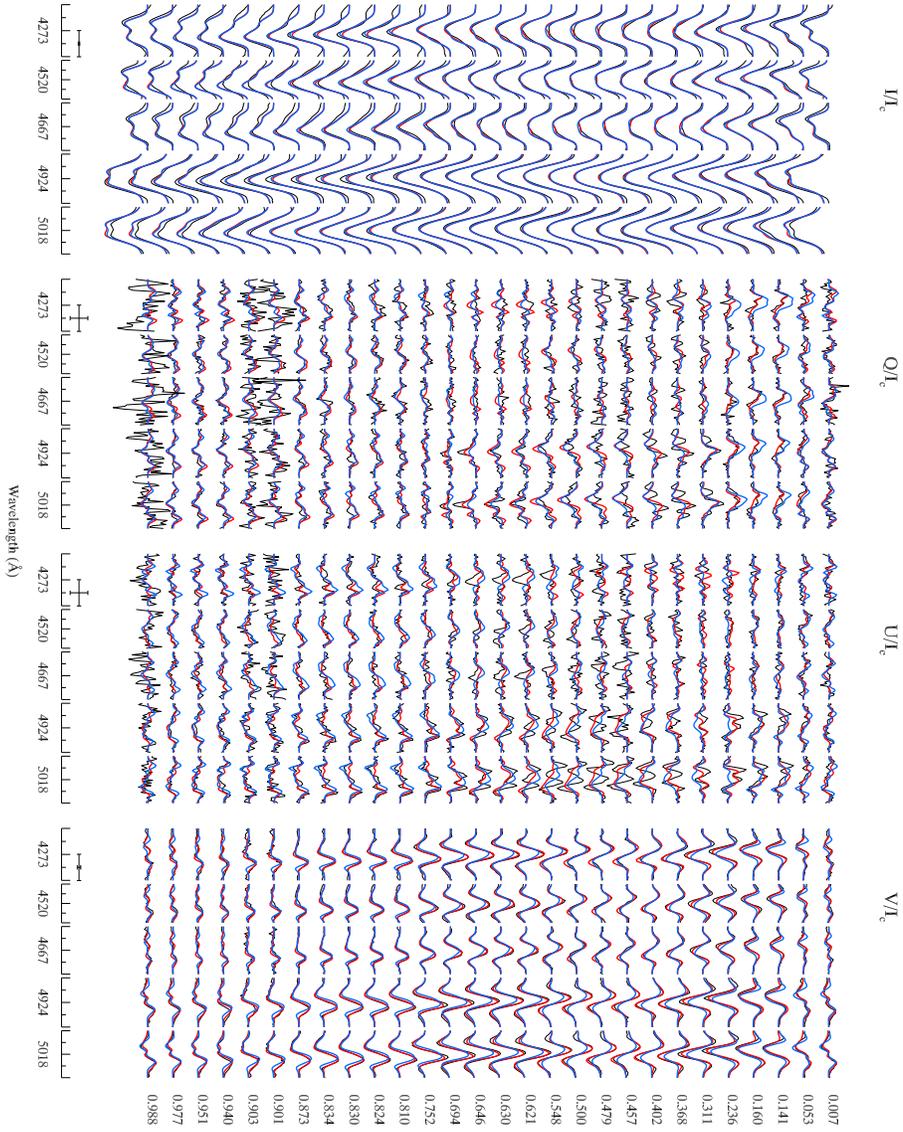}
   \caption{Comparison between observed (black lines) for the iron lines used in mapping and synthetic spectra for a dipolar geometry (solid curve, blue) and for a dipole + quadrupolar geometry (solid curve, red).}
\label{othergeofit}
\end{center}
\end{figure*}

As illustrated in Fig. \ref{othergeofit}, it can be seen that neither model is in good agreement with the profiles, in particular with the Stokes $QU$ profiles. We can therefore conclude even with the new data, that a simple field topology cannot describe the field structure of $\alpha^2$ CVn.

\section{Abundance Mapping}
We have also started to investigate which chemical elements are suitable for use in the reconstruction of surface abundance structures in $\alpha^2$ CVn.  The basis of line selection was to start with the lines used in the mapping by Kochukhov et al. (2002) and then to expand the list using the line list of Pyper (1969), Cohen (1970) and to a lesser extent Roby and Lambert (1990). This was combined with a visual inspection to eliminate lines from the list which were not clearly present in the spectrum, did not show variability or were heavily blended with other lines.  In addition lines which suffered from non-LTE affects were avoided. For the chemical abundance mapping only Stokes $IV$ profiles were to be used in the mapping, so lines did not have to exhibit linear polarisation signatures, but had to show clear variability in the absorption line.  The results of this investigation will be presented in a future paper (Silvester et al., in prep).

\section{Conclusions}
We have shown that the new ESPaDOnS/Narval data lead to a magnetic field topology consistent with the one found by using MuSiCoS data. This also provides direct observational evidence of the magnetic field stability in  $\alpha^2$ CVn over a decade timescale. We have also shown that the observations of iron lines in all Stokes parameters cannot be explained by a pure dipolar or quadrupolar model.  Further investigations into the importance of line selection when reconstructing the magnetic field topology of $\alpha^2$ CVn will be performed in paper two (Silvester et al., submitted). This will be followed by a study of all the abundance maps obtainable for $\alpha^2$ CVn with the new data set  (Silvester et al., in prep). 


\end{document}